\documentclass{iopconfser}

\usepackage{graphicx} 
\usepackage{ragged2e}
\usepackage{url}  
\usepackage{amsmath} 

\usepackage{hyperref}
\PassOptionsToPackage{hyperfootnotes=false,pdfpagelabels,linktocpage}{hyperref}
\hypersetup{%
	colorlinks=true, linktocpage=true, 
	breaklinks=true, pdfpagemode=UseNone, pageanchor=true,
	pdfpagemode=UseOutlines,%
	plainpages=false, bookmarksnumbered, bookmarksopen=true,
	bookmarksopenlevel=1,%
	hypertexnames=true, 
	urlcolor=blue, linkcolor=black, citecolor=blue,
}

\begin{document}

\title{URANOS - a novel voxel engine Neutron Transport Monte-Carlo Simulation}

\author{Markus K\"ohli$^{1}$, Martin Schr\"on$^{2}$, Steffen Zacharias$^{2}$ and Ulrich Schmidt$^{1}$}

\affil{$^1$Physikalisches Institut, Heidelberg University, Heidelberg, Germany}
\affil{$^2$Dep. Monitoring and Exploration Technologies, Helmholtz Centre for Environmental Research GmbH – UFZ, Leipzig, Germany}

\email{koehli@physi.uni-heidelberg.de}

\justify
\begin{abstract}
URANOS is a newly developed 3D neutron transport Monte-Carlo code from thermal to fast energy domains. It was originally developed for the CASCADE detector. The purpose of this simulation program is to provide a fast computational workflow and an intuitive graphical user interface (GUI) for small to medium-sized projects. It features a ray-casting algorithm based on a voxel engine. The simulation domain is defined layerwise, whereas the geometry is extruded from a pixel matrix of materials, identified by specific numbers. Input files are a stack of pictures, all other settings, including the configuration of predefined sources, can be adjusted via the GUI. 
The scattering kernel features the treatment of elastic and inelastic collisions, absorption and absorption-like processes like evaporation. Cross sections, energy distributions and angular distributions are taken from evaluated data bases. In order to simulate boron-lined detectors it also models the charged particle transport following the conversion by computing the energy loss in the boron and its consecutive layer. The electron track is then projected onto a readout unit by longitudinal and transversal diffusion.
URANOS is freely available and can be used to simulate the response function of boron-lined or epithermal neutron detectors, small-scale laboratory setups and especially transport studies of cosmic-ray induced environmental neutrons. It offers an easy accessibility and comparably simple interface capable of handling complex geometries. URANOS therefore offers possibilities to understand and simulate the neutron environment at instruments, which would otherwise require extensive modeling and training on dedicated packages. 

\end{abstract}

\section{Introduction}

Understanding neutron transport through Monte Carlo simulations is the backbone of instrument design and accurate interpretation of measurements. Over the past few decades, the Monte-Carlo code MCNP6~\cite{MCNP6} (Monte Carlo N-Particle) and its predecessor MCNPX~\cite{mcnpx} were often consulted to study the behavior of neutrons for detectors or instruments~\cite{NovelResponseMCNP,Tickner2001}. The conventional model accounts for various particles, including neutrons, photons, electrons and the their coupled transport, decreasing the computational efficiency and resulting in complex model structures. These impede adaptability and are especially challenging for novice users to operate. Versions until MCNP4 were written in Fortran, which to the present day shapes the user interaction and input logic where consistency was preferred rather than modernization. Users must still understand and adapt to the legacy design and rigid input format structure. The FLUKA~\cite{Fluka} (FLUktuierende KAskade) code is mostly oriented towards charged hadronic transport and nuclear and particle physics experiments. FLUKA is not directly linked against an evaluated data base, but operates on its own set of reprocessed and simplified mean values. 
GEANT4~\cite{Geant4} (GEometry ANd Tracking), originally developed for particle physics applications, provides a more modern Monte-Carlo tool, based on multithreaded C\texttt{++} code and OpenGL visualizations. GEANT4 especially excels in describing complex geometries. Since 2011, driven by requests from the European Spallation Source~\cite{TdrESS}, an increasing number of low-energy neutron calculation features were introduced. Meanwhile the software has advanced to a level where there is a good agreement with other codes like MCNP for fast neutrons~\cite{GEANTComparisonFast} as well as slow neutrons~\cite{GEANTComparisonDetector}, including Cosmic-Ray Neutron studies~\cite{Liu2021}.
Specific realizations with focus on instrument designs are McStas~\cite{Mcstas}, VITESS~\cite{Vitess} or RESTRAX~\cite{restrax}. NCrystal~\cite{NCrystal} is a tailored interfaces to existing simulations.
\noindent URANOS (Ultra Rapid Adaptable Neutron-Only Simulation)~\cite{URANOS2023} was developed as an alternative for the growing field of neutron research and instrument design. It was firstly developed to address open and recurring questions in environmental and hydrological sciences~\cite{myself,notMyself2,notMyself3,myself4,dazhi2019,Schattan2019,Weimar2020,Koehli2021,Badiee2021,Jakobi2021, gi-2021-18,Hubert2024}. As the model has evolved, it has been proven to be useful to simulate and drive the development of several type of neutron detectors~\cite{cascade2016,NovelNeutronDetectors,Koehli2022}. URANOS is computationally very efficient as it only accounts for the most relevant neutron interaction processes, namely elastic collisions, inelastic collisions, absorption, and evaporation. The main model features are: (A) tracking of particle histories from creation to detection, (B) detector representation as layers or geometric shapes, (C) voxel-based model extrusion and material setup based on color codes in ASCII matrices or bitmap images. After being used in detector development the main user base shifted to environmental sciences, which strongly influenced URANOS and its available settings.

\section{Concept and Interface}

\begin{figure}[t]
\centering
\includegraphics[width=0.95\textwidth]{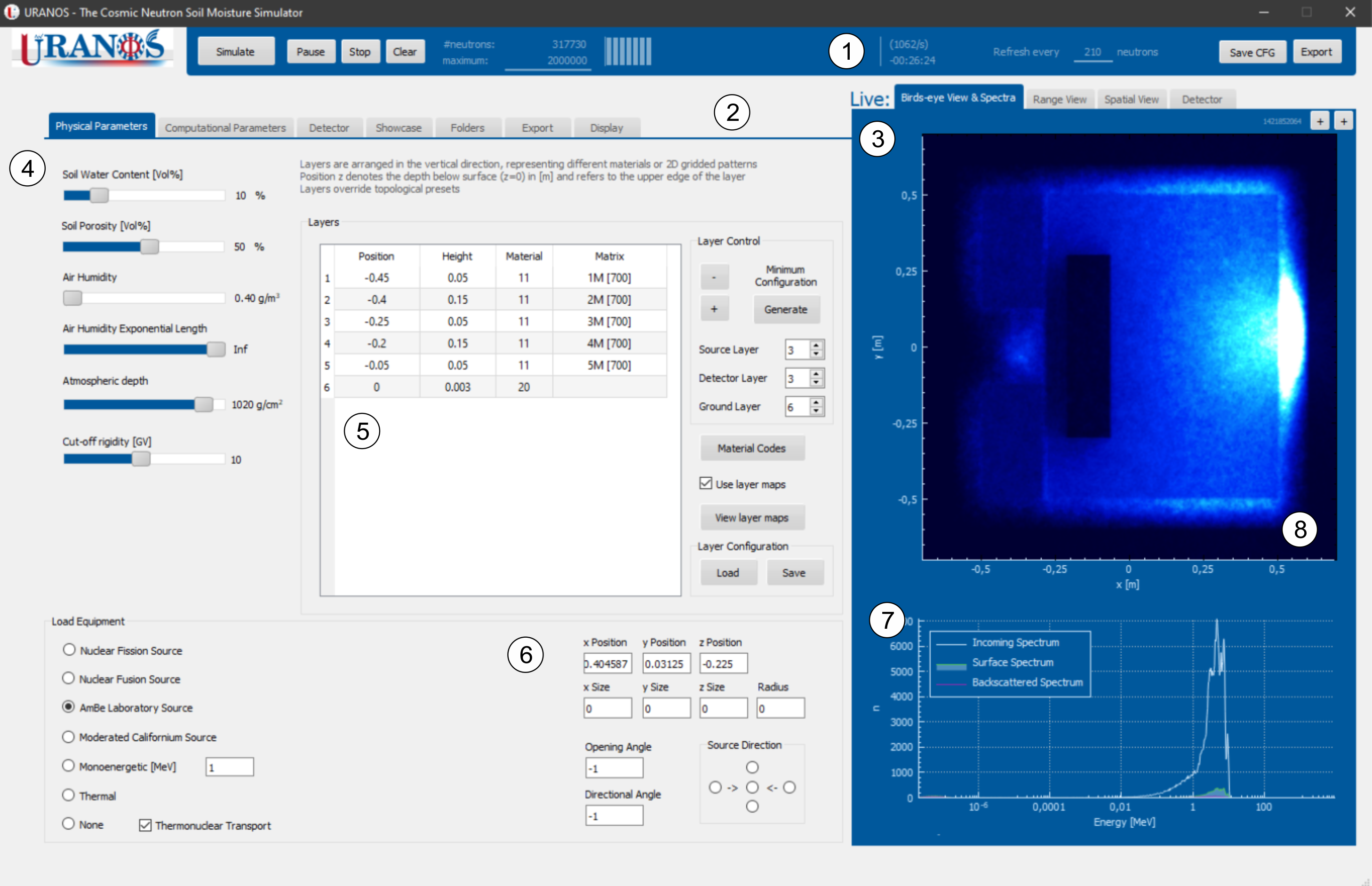}
\caption{URANOS main user interface for the calculation of a source inside a polyethylene box with (1) simulation control bar with status information, (2) general configuration tabs, (3) 'Live View' tabs for user feedback about the current run, (4) global environmental settings, (5) geometry stack with layers defined by voxels, (6) source definition (7), neutron energy spectrum above the ground layer and (8) cross section view of the neutron flux within the domain in the detector layer. }
\label{fig:interface}
\end{figure}

\noindent URANOS is designed as a Monte-Carlo tool which exclusively simulates contributions in an environment of neutron interactions. The standard calculation routine features a ray-casting algorithm for single neutron propagation and a voxel engine. The physics model follows the routines declared by the ENDF database standard and descriptions of implementations by OpenMC~\cite{openmcRef}. 
It features the treatment of elastic collisions in the thermal and epithermal regime, as well as inelastic collisions, absorption and emission processes such as evaporation, the delayed emission of MeV neutrons from excited nuclei. Cross sections, energy distributions and angular distributions were taken from the databases ENDF/B-VII.1~\cite{endfRef}, ENDF/B-VIII.0~\cite{endf8} and JENDL/HE-2007~\cite{jendlRef}. The entire software is developed in C\texttt{++} and linked against CERN's ROOT~\cite{root}, whereas the GUI uses the QT cross-platform framework. The graphical user interface, see Fig.~\ref{fig:interface}, offers features specifically tailored to quickly solve general tasks in a relatively simple way. But it also particularly targets research questions in complicated contexts that a voxel geometry can represent.

\section{Usage and workflow}

URANOS can be executed from the command line using \texttt{'UranosGUI.exe tnw'} for the GUI version or  \texttt{'UranosGUI.exe tnw config path/to/uranos.cfg'} for the command-line version.

\noindent The graphical user interface provides utilities for all settings for adjusting the simulation. The main window with a selection of highlighted features (1)-(8) is displayed in Fig.~\ref{fig:interface}. The main user space is split into the control bar (1), tabs for the settings (2) and for the display of live information during the runtime (3). The sliders control general features of air and soil (4), which can be used as materials in the geometry stack (5). Available sources cover fixed energies, thermal neutrons, a partially moderated spectrum and standard laboratory fusion and fission sources (6). After starting the simulation, the neutron distribution within the domain can be viewed in the visualization tab (7) and (8). 
\noindent URANOS uses analytical geometry definitions and voxels as introduced earlier. The following top-down structure is applied to describe the simulation:
$\mathrm{geometry}\rightarrow \mathrm{layer}\rightarrow \mathrm{voxel mesh}\rightarrow \mathrm{material}\rightarrow \mathrm{isotope}$.
\noindent Each layer of the stack is either entirely filled by a material or subdivided into several sections using a two-dimensional matrix, which will be extruded to voxels filled with predefined materials. A material is a specific composition of isotopes with atomic weight and density. Most definitions are taken from~\cite{MatCompendium}. 
\noindent Neutrons are then propagated through the domain via ray casting, a technique where each event is calculated sequentially for modeling the particle tracks. URANOS can score various quantities of possible outputs, for example, the flux density above the ground. Neutrons in general can be scored in three different ways: one layer can be defined as the 'detector layer', a virtual entity which can record any particle of chosen characteristics to pass through. The detector layer spans the full width of the simulation domain and is placed at a fixed height. Secondly, a virtual detector with limited spatial extension can be placed within the detector layer, which can represent for example a counter tube. Specific 'detector' materials applied to voxels can additionally score neutrons like the virtual detector. Output options are either only detector hits, partial tracks within a material or the full track history. The virtual detectors can emulate an instrument by directly applying an energy and angle-dependent response function to it.

\begin{figure}[t]
\centering
\includegraphics[width=\textwidth]{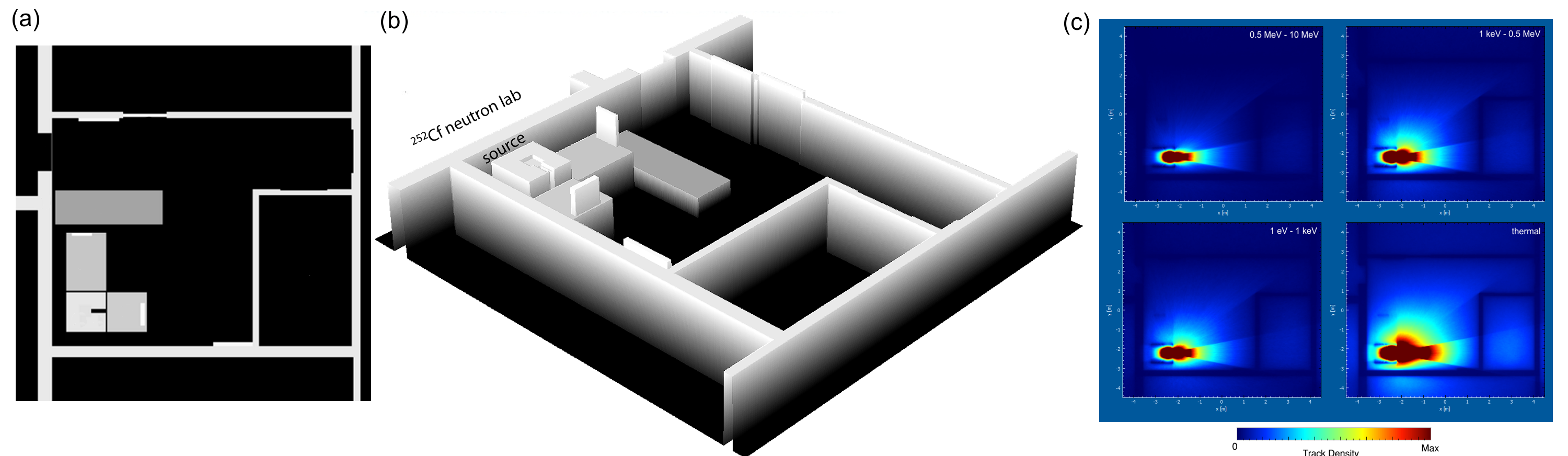}
\caption{URANOS example for the calculation of a neutron flux distribution within a laboratory with a $^{252}$Cf source: (a) stacked input geometry with grayscale codes for different materials, (b) three-dimensional extrusion of the layers and (c) result of the calculation for different energy windows from the MeV range over fast and epithermal neutrons to the thermal regime.  }
\label{fig:CfExample}
\end{figure}

\noindent A simple example for the calculation of the neutron flux from a laboratory source is shown in Fig.~\ref{fig:CfExample}. Here, the moderated californium source sits in the lower left part of the room within a biological shield including two beamports with shutters. The user defines the geometry of the room with the floor, walls, windows, doors and experimental stations. The beam of the source is divergent and the angular distribution is highly inhomogeneous within the experimental area. After running the simulation within the domain the user obtains a better understanding for the neutron distribution in the room itself, can estimate the local beam divergence for further setups or estimate the background. The computational speed for this example is approximately 300 neutrons per CPU GHz per core.    

\begin{figure}[t]
\centering
\includegraphics[width=\textwidth]{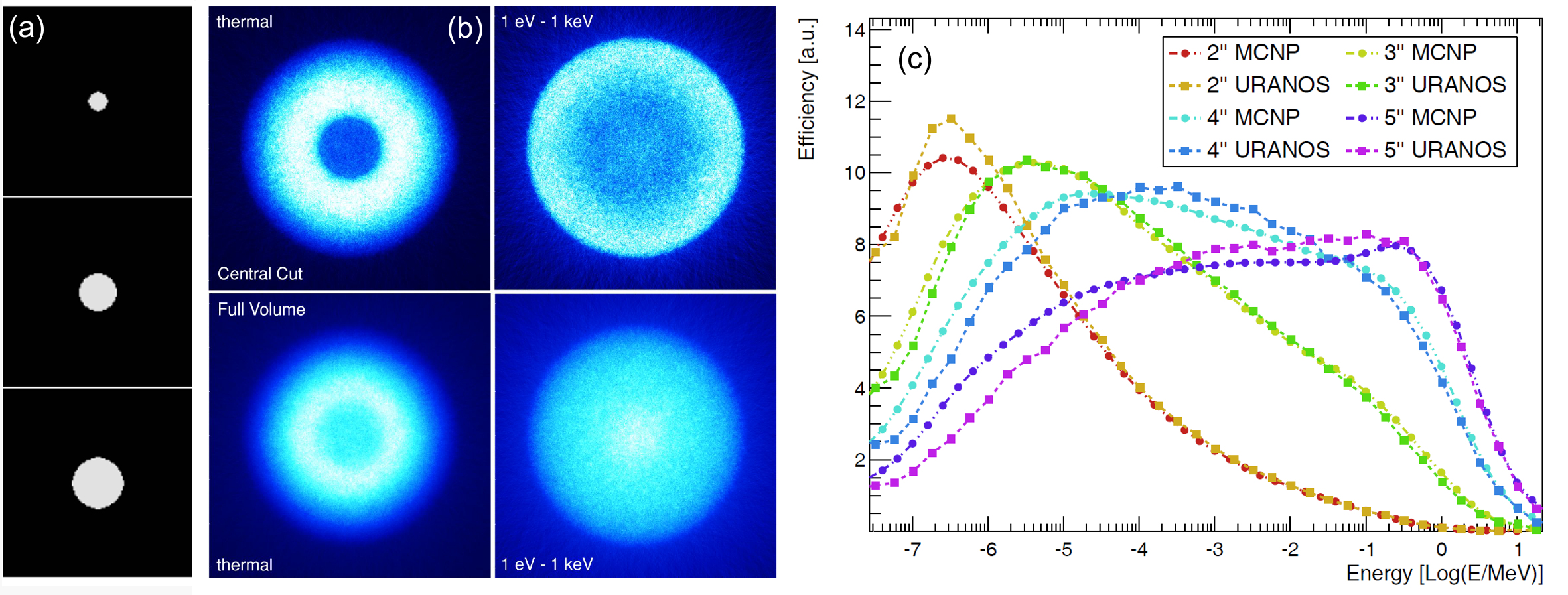}
\caption{URANOS example for the calculation of response functions of Bonner Spheres: (a) part of the stacked input layer geometry, (b) cross section of the neutron flux through the spheres (top row) or integrated over the entire volume (lower row) for different energy ranges and (c) result of the response function calculation and comparison to~\cite{Mares3He}.  }
\label{fig:BonnerExample}
\end{figure}

\noindent URANOS can also be used for the calculation of detector response functions. As an example Bonner Spheres~\cite{Bonner} with spherical helium-3 counters are shown in Fig.~\ref{fig:BonnerExample}. This detector array is used to unfold neutron spectra and various studies were carried out for simulating its sensitivity, the detector model in this example is from~\cite{Mares3He}. The graphical representation of the neutron tracks in blue serves only illustrative purposes. The response functions can be obtained by a predefined batch run function sweeping over the entire energy range. Although the spheres were discretized into voxels, the results compare well with previous MCNP calculations. 

\section{Conclusion}

URANOS is a novel neutron-only Monte-Carlo tool based on C\texttt{++} with an optional graphical user interface. It uniquely features a modeling geometry using the concept of voxels, three-dimensional pixels, which can be stacked in layers and extruded from grayscale images or ASCII matrices, not requiring complex steering files. Several built-in neutron spectra for typical laboratory sources provide options to directly start calculations with the input geometry. It allows to record the neutron flux by a domain-wide scoring layer and a virtual detector with defined instrument characteristics. These built-in scoring methods short-cut the data analysis toolchain for modeling without requiring programming skills. URANOS features the computation of all relevant neutron interactions below 20\,MeV. For higher energies URANOS uses an effective cascade model which reproduces the empirically know flux by adjustable parameters. Comparisons show a close agreement of the URANOS model with other simulation toolkits like MCNP. 
URANOS is now available for all users and regularly maintained to address the growing needs of users.

\section*{Competing Interests}
M. K\"ohli holds a CEO position at StyX Neutronica GmbH, Germany.

\section*{Code Availability}
URANOS is made available for Windows and Unix platforms on GitLab  \url{https://gitlab.com/mkoehli/uranos}, including a collection of examples, Wiki pages and use guides. URANOS v1.0 has been released under DOI: 10.7910/DVN/THPNZW. Furthermore, the code (v1.0) has been released under DOI: 10.5281/zenodo.6578668. The current model of URANOS (v1.27) is also available from the project website \url{https://www.physi.uni-heidelberg.de/Forschung/ANP/Cascade/URANOS/}.

\section*{Acknowledgements}

URANOS was developed within several projects. Initial funding was provided by the projects ‘Neutron Detectors for the MIEZE method’ and ’Forschung und Entwicklung hochaufl\"osender Neutronendetektoren’, funded by the German Federal Ministry for Research and Education (BMBF), grant identifier: 05K10VHA and 05K16PD1 and by the DFG (German Research Foundation) research unit FOR 2694 Cosmic Sense via the project 357874777.

\bibliographystyle{unsrt}
\bibliography{literature.bib}

@Article{endfRef,
  author   = {M.B. Chadwick and others},
  title    = {{ENDF/B-VII.1} Nuclear Data for Science and Technology: Cross Sections, Covariances, Fission Product Yields and Decay Data},
  doi      = {10.1016/j.nds.2011.11.002},
  issn     = {0090-3752},
  note     = {Spec. Issue ENDF/B-VII.1 Library},
  number   = {12},
  pages    = {2887--2996},
  url      = {https://www.sciencedirect.com/science/article/pii/S009037521100113X},
  volume   = {112},
  fjournal = {Nuclear Data Sheets},
  journal  = {Nucl. Data Sheets},
  year     = {2011},
}

@Article{endf8,
  author   = {D.A. Brown and others},
  title    = {{ENDF/B-VIII.0}: The 8$^{\text{th}}$ Major Release of the Nuclear Reaction Data Library with {CIELO}-project Cross Sections, New Standards and Thermal Scattering Data},
  doi      = {10.1016/j.nds.2018.02.001},
  issn     = {0090-3752},
  note     = {Spec. Issue Nucl. React. Data},
  pages    = {1--142},
  url      = {https://www.sciencedirect.com/science/article/pii/S0090375218300206},
  volume   = {148},
  fjournal = {Nuclear Data Sheets},
  journal  = {Nucl. Data Sheets},
  year     = {2018},
}

@Article{jendlRef,
  author   = {Shibata, K. and others},
  title    = {{JENDL}-4.0: A New Library for Nuclear Science and Engineering},
  doi      = {10.1080/18811248.2011.9711675},
  number   = {1},
  pages    = {1--30},
  url      = {https://www.tandfonline.com/doi/abs/10.1080/18811248.2011.9711675},
  volume   = {48},
  fjournal = {Journal of Nuclear Science and Technology},
  journal  = {J. Nucl. Sci. Technol.},
  year     = {2011},
}

@Article{myself,
  author    = {K{\"o}hli, M. and others},
  title     = {Footprint characteristics revised for field-scale soil moisture monitoring with cosmic-ray neutrons},
  doi       = {10.1002/2015WR017169},
  issn      = {1944-7973},
  number    = {7},
  pages     = {5772--5790},
  url       = {https://onlinelibrary.wiley.com/doi/10.1002/2015WR017169/full},
  volume    = {51},
  fjournal  = {Water Resources Research},
  journal   = {Water Resour. Res.},
  shorthand = {KS2015},
  year      = {2015},
}

@Article{cascade2016,
  author    = {M. K{\"o}hli and others},
  title     = {Efficiency and spatial resolution of the {CASCADE} thermal neutron detector},
  doi       = {10.1016/j.nima.2016.05.014},
  issn      = {0168-9002},
  pages     = {242--249},
  url       = {https://www.sciencedirect.com/science/article/pii/S0168900216303722},
  volume    = {828},
  fjournal  = {Nuclear Instruments and Methods in Physics Research Section A: Accelerators, Spectrometers, Detectors and Associated Equipment},
  journal   = {Nucl. Instrum. Methods Phys. Res., Sect. A},
  shorthand = {K2016},
  year      = {2016},
}

@Article{myself4,
  author    = {M. K{\"o}hli and others},
  title     = {Response functions for detectors in cosmic ray neutron sensing},
  doi       = {10.1016/j.nima.2018.06.052},
  issn      = {0168-9002},
  pages     = {184--189},
  url       = {https://www.sciencedirect.com/science/article/pii/S0168900218307745},
  volume    = {902},
  fjournal  = {Nuclear Instruments and Methods in Physics Research Section A: Accelerators, Spectrometers, Detectors and Associated Equipment},
  journal   = {Nucl. Instrum. Methods Phys. Res., Sect. A},
  shorthand = {KS2018},
  year      = {2018},
}

@Article{NovelNeutronDetectors,
  author    = {M. K{\"o}hli and others},
  title     = {Novel neutron detectors based on the time projection method},
  doi       = {10.1016/j.physb.2018.03.026},
  issn      = {0921-4526},
  note      = {The 11$^{\text{th}}$ Int. Conf. Neutron Scatter. ({ICNS} 2017)},
  pages     = {517--522},
  url       = {https://www.sciencedirect.com/science/article/pii/S0921452618302187},
  volume    = {551},
  fjournal  = {Physica B: Condensed Matter},
  journal   = {Physica B},
  shorthand = {K2018},
  year      = {2018},
}

@Article{notMyself2,
  author    = {Schr{\"o}n, M. and others},
  title     = {Improving Calibration and Validation of Cosmic-Ray Neutron Sensors in the Light of Spatial Sensitivity},
  doi       = {10.5194/hess-21-5009-2017},
  number    = {10},
  pages     = {5009--5030},
  url       = {https://www.hydrol-earth-syst-sci.net/21/5009/2017/},
  volume    = {21},
  fjournal  = {Hydrology and Earth System Sciences},
  journal   = {Hydrol. Earth Syst. Sc.},
  shorthand = {SK2017},
  year      = {2017},
}

@Article{notMyself3,
  author    = {Schr\"{o}n, M. and others},
  title     = {{Intercomparison of cosmic-ray neutron sensors and water balance monitoring in an urban environment}},
  doi       = {10.5194/gi-7-83-2018},
  number    = {1},
  pages     = {83--99},
  volume    = {7},
  fjournal  = {Geoscientific Instrumentation, Methods and Data Systems},
  journal   = {Geosci. Instrum. Methods Data Syst.},
  shorthand = {SK2017b},
  year      = {2018},
}

@Article{openmcRef,
  author   = {P.K. Romano and B. Forget},
  title    = {The {OpenMC} {Monte Carlo} particle transport code},
  doi      = {10.1016/j.anucene.2012.06.040},
  issn     = {0306-4549},
  pages    = {274--281},
  url      = {https://www.sciencedirect.com/science/article/pii/S0306454912003283},
  volume   = {51},
  fjournal = {Annals of Nuclear Energy},
  journal  = {Ann. Nucl. Energy},
  year     = {2013},
}

@Article{Fluka,
  author   = {G. Battistoni and others},
  title    = {Overview of the {FLUKA} code},
  doi      = {10.1016/j.anucene.2014.11.007},
  issn     = {0306-4549},
  note     = {Joint International Conference on Supercomputing in Nuclear Applications and Monte Carlo 2013},
  pages    = {10--18},
  url      = {https://www.sciencedirect.com/science/article/pii/S0306454914005878},
  volume   = {82},
  fjournal = {Annals of Nuclear Energy},
  journal  = {Ann. Nucl. Energy},
  year     = {2015},
}

@Article{mcnpx,
  author  = {L.S. Waters and others},
  title   = {The {MCNPX} {Monte} {Carlo} Radiation Transport Code},
  doi     = {10.1063/1.2720459},
  number  = {1},
  pages   = {81--90},
  url     = {https://aip.scitation.org/doi/abs/10.1063/1.2720459},
  volume  = {896},
  journal = {{AIP} Conf. Proc.},
  year    = {2007},
}

@Article{MCNP6,
  author    = {T. Goorley and others},
  title     = {Initial {MCNP6} Release Overview},
  doi       = {10.13182/NT11-135},
  number    = {3},
  pages     = {298--315},
  url       = {http://www.ans.org/pubs/journals/nt/a_15346},
  volume    = {180},
  fjournal  = {Nuclear Technology},
  journal   = {Nucl. Technol.},
  publisher = {Taylor \& Francis},
  year      = {2012},
}

@Article{GEANTComparisonFast,
  author  = {A.N. Solovyev and others},
  title   = {Comparative analysis of {MCNPX} and {GEANT4} codes for fast-neutron radiation treatment planning},
  doi     = {10.1016/j.nucet.2015.11.004},
  issn    = {2452-3038},
  number  = {1},
  pages   = {14--19},
  url     = {https://www.sciencedirect.com/science/article/pii/S2452303815000059},
  volume  = {1},
  journal = {Nucl. Energy Technol.},
  year    = {2015},
}

@Article{GEANTComparisonDetector,
  author   = {B.M. {van der Ende} and others},
  title    = {Use of {GEANT4} vs. {MCNPX} for the characterization of a boron-lined neutron detector},
  doi      = {10.1016/j.nima.2016.02.082},
  issn     = {0168-9002},
  pages    = {40--47},
  url      = {https://www.sciencedirect.com/science/article/pii/S016890021630002X},
  volume   = {820},
  fjournal = {Nuclear Instruments and Methods in Physics Research Section A: Accelerators, Spectrometers, Detectors and Associated Equipment},
  journal  = {Nucl. Instrum. Methods Phys. Res., Sect. A},
  year     = {2016},
}

@Article{Geant4,
  author   = {S. Agostinelli and others},
  title    = {{GEANT4} - a simulation toolkit},
  doi      = {10.1016/S0168-9002(03)01368-8},
  issn     = {0168-9002},
  number   = {3},
  pages    = {250--303},
  url      = {https://www.sciencedirect.com/science/article/pii/S0168900203013688},
  volume   = {506},
  fjournal = {Nuclear Instruments and Methods in Physics Research Section A: Accelerators, Spectrometers, Detectors and Associated Equipment},
  journal  = {Nucl. Instrum. Methods Phys. Res., Sect. A},
  year     = {2003},
}

@Book{TdrESS,
  author    = {Peggs, S. and others},
  title     = {{European Spallation Source Technical Design Report}},
  isbn      = {978-91-980173-2-8},
  note      = {{ESS}-2013-001},
  pagetotal = {690},
  publisher = {ESS},  
  url       = {http://docdb01.esss.lu.se/DocDB/0002/000274/006/TDR_final_130423_print_ch1.pdf},
  address   = {Lundt},
  year      = {2013},
}

@InProceedings{root,
  author    = {Brun, R. and Rademakers, F.},
  booktitle = {Proc. {AIHENP}’96 Workshop, Lausanne},
  title     = {{ROOT} - An Object Oriented Data Analysis Framework},
  doi       = {10.1016/S0168-9002(97)00048-X},
  note      = {see also \texttt{https://root.cern.ch/}},
  pages     = {81--86},
  volume    = {389},
  fjournal  = {Nuclear Instruments and Methods in Physics Research Section A: Accelerators, Spectrometers, Detectors and Associated Equipment},
  journal   = {Nucl. Instrum. Methods Phys. Res., Sect. A},
  year      = {1997},
}

@TechReport{MatCompendium,
  author       = {{McConn~Jr.}, R.J. and others},
  institution  = {Pacific Northwest National Laboratory},
  title        = {Compendium of Material Composition Data for Radiation Transport Modeling},
  number       = {PNNL-15870 Rev. 1},
  pagetotal    = {375},
  url          = {https://www.pnnl.gov/main/publications/external/technical_reports/pnnl-15870rev1.pdf},
  address      = {Richland, Washington 99352},
  reportnumber = {PNNL-15870 Rev. 1},
  year         = {2011},
}

@Article{Bonner,
  author   = {R.L. Bramblett and others},
  title    = {A new type of neutron spectrometer},
  doi      = {10.1016/0029-554X(60)90043-4},
  issn     = {0029-554X},
  number   = {1},
  pages    = {1--12},
  url      = {https://www.sciencedirect.com/science/article/pii/0029554X60900434},
  volume   = {9},
  fjournal = {Nuclear Instruments and Methods},
  journal  = {Nucl. Instrum. Methods},
  year     = {1960},
}

@Article{Mares3He,
  author   = {V. Mares and others},
  title    = {Calculated neutron response of a {B}onner {S}phere {S}pectrometer with $^{3}${He} counter},
  doi      = {10.1016/0168-9002(91)90210-H},
  issn     = {0168-9002},
  number   = {2},
  pages    = {398--412},
  url      = {https://www.sciencedirect.com/science/article/pii/016890029190210H},
  volume   = {307},
  fjournal = {Nuclear Instruments and Methods in Physics Research Section A: Accelerators, Spectrometers, Detectors and Associated Equipment},
  journal  = {Nucl. Instrum. Methods Phys. Res., Sect. A},
  year     = {1991},
}

@Article{NovelResponseMCNP,
  author   = {A.W. Decker and others},
  title    = {Novel {B}onner {S}phere {S}pectrometer Response Functions Using {MCNP6}},
  doi      = {10.1109/TNS.2015.2416652},
  issn     = {0018-9499},
  number   = {4},
  pages    = {1689--1694},
  url      = {https://ieeexplore.ieee.org/document/7104175/},
  volume   = {62},
  fjournal = {IEEE Transactions on Nuclear Science},
  journal  = {IEEE Trans. Nucl. Sci.},
  year     = {2015},
}

@Article{dazhi2019,
  author   = {D. Li and others},
  title    = {Can Drip Irrigation be Scheduled with Cosmic-Ray Neutron Sensing?},
  doi      = {10.2136/vzj2019.05.0053},
  number   = {1},
  pages    = {190053},
  url      = {https://dl.sciencesocieties.org/publications/vzj/abstracts/18/1/190053},
  volume   = {18},
  fjournal = {Vadose Zone Journal},
  journal  = {Vadose Zone J.},
  year     = {2019},
}

@Article{Schattan2019,
  author   = {Schattan, P. and others},
  title    = {Sensing Area-Average Snow Water Equivalent with Cosmic-Ray Neutrons: The Influence of Fractional Snow Cover},
  doi      = {10.1029/2019WR025647},
  number   = {12},
  pages    = {10796--10812},
  url      = {https://agupubs.onlinelibrary.wiley.com/doi/abs/10.1029/2019WR025647},
  volume   = {55},
  fjournal = {Water Resources Research},
  journal  = {Water Resour. Res.},
  year     = {2019},
}

@Article{Weimar2020,
  author  = {Weimar, J. and others},
  title   = {Large-Scale Boron-Lined Neutron Detection Systems as a $^3${H}e Alternative for Cosmic Ray Neutron Sensing},
  doi     = {10.3389/frwa.2020.00016},
  pages   = {16},
  url     = {https://www.frontiersin.org/article/10.3389/frwa.2020.00016},
  volume  = {2},
  journal = {Front. Water},
  year    = {2020},
}

@Article{Koehli2021,
  author  = {K\"ohli, M. and others},
  title   = {Soil Moisture and Air Humidity Dependence of the Above-Ground Cosmic-Ray Neutron Intensity},
  doi     = {10.3389/frwa.2020.544847},
  pages   = {15},
  url     = {https://doi.org/10.3389/frwa.2020.544847},
  volume  = {2},
  journal = {Front. Water},
  year    = {2021},
}

@Article{Jakobi2021,
  author    = {J. Jakobi and others},
  title     = {The Footprint Characteristics of Cosmic Ray Thermal Neutrons},
  doi       = {10.1029/2021gl094281},
  number    = {15},
  pages     = {e2021GL094281},
  url       = {https://doi.org/10.1029/2021gl094281},
  volume    = {48},
  fjournal  = {Geophysical Research Letters},
  journal   = {Geophys. Res. Lett.},
  publisher = {American Geophysical Union ({AGU})},
  year      = {2021},
}

@Article{gi-2021-18,
  author   = {Francke, T. and others},
  title    = {Assessing the feasibility of a directional cosmic-ray neutron sensing sensor for estimating soil moisture},
  doi      = {10.5194/gi-11-75-2022},
  number   = {1},
  pages    = {75--92},
  url      = {https://gi.copernicus.org/articles/11/75/2022/},
  volume   = {11},
  fjournal = {Geoscientific Instrumentation, Methods and Data Systems},
  journal  = {Geosci. Instrum. Methods Data Syst.},
  year     = {2022},
}

@Article{Badiee2021,
  author    = {A. Badiee and others},
  title     = {Using Additional Moderator to Control the Footprint of a {COSMOS} Rover for Soil Moisture Measurement},
  doi       = {10.1029/2020wr028478},
  number    = {6},
  pages     = {e2020WR028478},
  url       = {https://doi.org/10.1029/2020wr028478},
  volume    = {57},
  fjournal  = {Water Resources Research},
  journal   = {Water Resour. Res.},
  publisher = {American Geophysical Union ({AGU})},
  year      = {2021},
}

@Article{Koehli2022,
  author   = {M. K\"ohli and J.-P. Schmoldt},
  title    = {Feasibility of {UXO} detection via pulsed neutron-neutron logging},
  doi      = {10.1016/j.apradiso.2022.110403},
  issn     = {0969-8043},
  pages    = {110403},
  url      = {https://www.sciencedirect.com/science/article/pii/S0969804322002895},
  volume   = {188},
  fjournal = {Applied Radiation and Isotopes},
  journal  = {Appl. Radiat. Isotopes},
  year     = {2022},
}

@Article{Liu2021,
  author   = {H. Liu and others},
  title    = {Cosmic-ray neutron fluxes and spectra at different altitudes based on {Monte} {Carlo} simulations},
  doi      = {10.1016/j.apradiso.2021.109800},
  issn     = {0969-8043},
  pages    = {109800},
  url      = {https://www.sciencedirect.com/science/article/pii/S0969804321002037},
  volume   = {175},
  fjournal = {Applied Radiation and Isotopes},
  journal  = {Appl. Radiat. Isotopes},
  year     = {2021},
}

@Article{Vitess,
  author    = {D. Wechsler and others},
  title     = {{VITESS}: Virtual instrumentation tool for pulsed and continuous sources},
  doi       = {10.1080/10448630008233764},
  number    = {4},
  pages     = {25--28},
  url       = {https://doi.org/10.1080/10448630008233764},
  volume    = {11},
  journal   = {Neutron News},
  publisher = {Informa {UK} Limited},
  year      = {2000},
}

\end{document}